\documentclass[a4paper,conference]{IEEEtran}
\IEEEoverridecommandlockouts
\usepackage{cite}
\usepackage{amsmath,amssymb,amsfonts}
\usepackage{algorithmic}
\usepackage{graphicx}
\usepackage{textcomp}
\usepackage{xcolor}

\def\BibTeX{{\rm B\kern-.05em{\sc i\kern-.025em b}\kern-.08em
    T\kern-.1667em\lower.7ex\hbox{E}\kern-.125emX}}
\begin{document}

\title{Overview of Fault Tolerant Techniques in Underwater Sensor Networks}

\author{\IEEEauthorblockN{Lauri Vihman}
\IEEEauthorblockA{\textit{Department of Computer Systems} \\
\textit{Tallinn University of Technology}\\
Tallinn, Estonia \\
lauri.vihman@taltech.ee}
\and
\IEEEauthorblockN{Maarja Kruusmaa}
\IEEEauthorblockA{\textit{Department of Computer Systems} \\
\textit{Tallinn University of Technology}\\
Tallinn, Estonia \\
maarja.kruusmaa@taltech.ee}
\and
\IEEEauthorblockN{Jaan Raik}
\IEEEauthorblockA{\textit{Department of Computer Systems} \\
\textit{Tallinn University of Technology}\\
Tallinn, Estonia \\
jaan.raik@taltech.ee}
}

\showboxbreadth=50 
\showboxdepth=50
\maketitle
% DDECS

\begin{abstract}
Sensor networks provide services to a broad range of applications ranging from intelligence service surveillance to weather forecasting. Most of the sensor networks are 
terrestrial, however much of our planet is covered by water and  Underwater Sensor Networks (USN) are an emerging research area. 
One of the unavoidable increasing challenge for modern technology is tolerating faults - accepting that hardware is imperfect and cope with it. Fault tolerance may have more 
impact underwater than in terrestrial environment as terrestrial environment is more forgiving, reaching the malfunctioning devices for replacement underwater is harder and may be 
more costly. Current paper is the first to investigate fault tolerance, particularly cross layer fault tolerance, in USN-s.
\end{abstract}

\section{Introduction}
\label{sect:intro}
\begin{figure}[htbp]
\centerline{\includegraphics[width=\columnwidth]{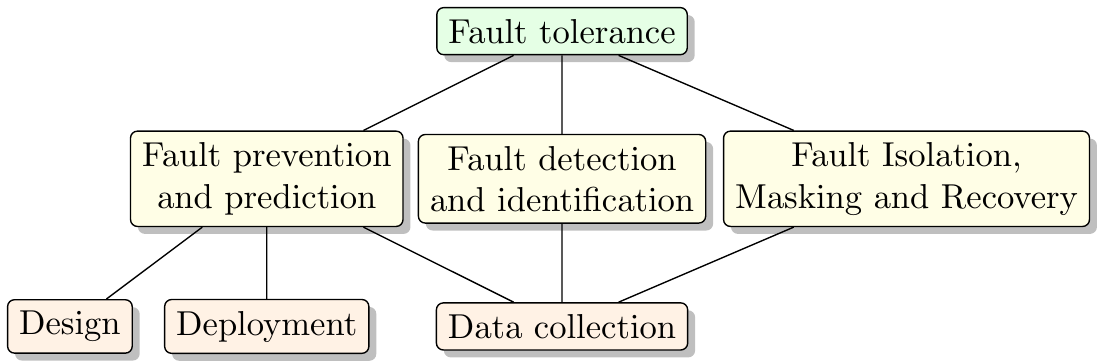}}
% \textwidth
\caption{Fault tolerance tasks in USN-s}
\label{fig:tasks}
\end{figure}

% ======================================INTRODUCTION===============================================
In current paper applications, practices, and central issues on Fault Tolerant Underwater Sensor Networks (USN-s) from previous research works are discussed. Because the global 
community has not yet put much effort on research of Fault Tolerance of USN-s, the criteria is expanded and papers covering only some parts of the topic are also taken into 
account. Many of the technologies, approaches and tools may possibly be adapted for use in USN-s. Fig.~\ref{fig:tasks}. shows the tasks of Fault Tolerance applicable in 
USN-s and how they affect each other. While design and initial deployment of USN-s contribute to Fault Prevention and Prediction abilities, data collecting techniques at the 
runtime contribute also to Fault Detection and Fault Recovery stages of the system, all of which are going to be discussed in current paper.

The rest of this paper is organized as follows: Further in current section methodology of selecting papers is explained, possible fault sources and categorization of 
techniques is discussed. Following sections are divided like shown on Fig.~\ref{fig:tasks}. In~\ref{sect:prevention} Fault Prevention and Prediction section design, deployment, 
data collection and testing frameworks are overviewed. In~\ref{sect:detection} Fault Detection and Identification section techniques for fulfilling those tasks are discussed. 
In~\ref{sect:recovery} Fault Masking and Recovery section relevant techniques are overviewed. In~\ref{sect:open} open research issues are discussed and finally 
in~\ref{sect:conclusion} current work is concluded.
% Devices are imperfect and Resilience and Sensor networks are in early stages of evolution Although when the technology matures, it will be time to regard more to the largest 
% and least known environment on our planet - underwater.  There have not yet been enough interest in USN % got high research interest lately, be it  emerging 
% Current paper investigates state of the art, practicies and central issues in fault tolerance of underwater sensor networks. , the criteria was expanded and papers covering only 
% some parts of the topic also taken into account. We want to know what's going on on fault tolerance studies for underwater stuff.

\subsection{Methodology}
\label{sect:method}
\begin{figure}[htbp]
\centerline{\includegraphics[width=\columnwidth]{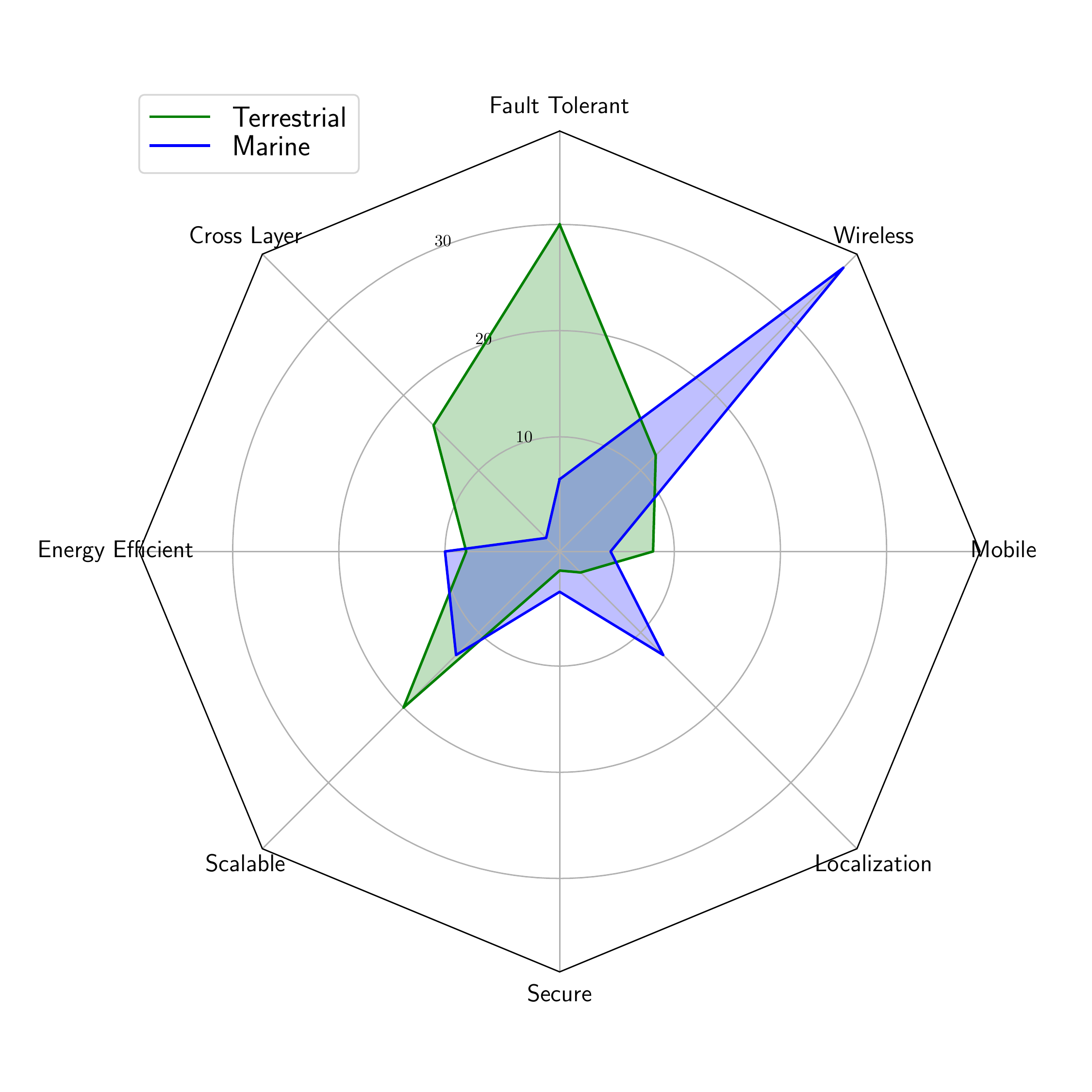}}
\caption{Number of analyzed papers touching specific categories}
\label{fig:radar}
\end{figure}

In order to obtain relevant sample in field of fault tolerant USN-s IEEE Explore, Google Scholar, Sciencedirect and Espacenet online environments were used with following search 
keywords in different combinations: ``underwater'',  ``sensor network'', ``internet of things'', ``resilient'', ``tolerant'', ``fault management'', ``cross-layer''. Top papers 
were selected by relevance order offered by environments' algorithms and sources on the topic found from those papers. Other citations of those sources were searched and more 
papers found this way. Related articles offered by IEEE Explore and Sciencedirect algorithms were also taken into account. Collected papers were next analyzed, classified and 
divided into marine and terrestrial categories, and the number  of papers managing specific areas of research are shown on radar diagram Fig.~\ref{fig:radar}. It should be noted 
that in the context of Fig.~\ref{fig:radar} meaning of ``localization'' is location detection in space and meaning of ``mobile'' is capacity of movement.
It can be seen from Fig.~\ref{fig:radar} that large share of marine research interest from the found papers has been drawn to underwater wireless communication while some on 
underwater fault tolerance techniques and almost none to underwater cross-layer fault tolerance. Underwater energy-efficiency and scalability are more covered areas than underwater 
vehicles (mobility) and security. Terrestrial papers were, according to initial search criteria, more concerned on fault-tolerance, including cross-layer fault tolerance, and less 
on energy-efficiency or security. 

High research effort on marine wireless networking in Fig.~\ref{fig:radar} conforms the claim that current pace of research on Internet of Underwater Things is slow due to 
the challenges arising from the uniqueness of underwater wireless sensor networks ~\cite{Kao2017}. Specifically, the main challenges for Internet of Underwater Things are the 
differences between Underwater Wireless Sensor Networks and Terrestrial Wireless Sensor Networks ~\cite{Kao2017}.

\subsection{Sources of Faults}
\label{sect:sources}
A fault is defined~\cite{Khosrow-PourD.B.A.2018} as an underlying defect of a system that leads to an error. 
Error is a faulty system state, which may lead to failure, and failure is an error that affects system functionality.
Faults may occur in different components and layers of systems for different reasons. The only type of fault possible in software is a design fault introduced during the software 
development i.e bug~\cite{Wilfredo2000}. The software bugs can be addressed separately and will not be covered further in current paper. 

Fault sources can be categorized by components where they occur. In sensor networks they  can occur in  sensor nodes, network and the data sink~\cite{Jaynes2013}. 
Sensor networks share common failure issues with traditional networks as well as introduce node failures as new fault sources~\cite{Paradis2007}. 

USN-s introduce additionally faults caused by environment conditions such as pressure, currents, underwater obstacles, etc. Those conditions may cause physical 
damage that may result in failures as well as obstruct system functionality. For instance in underwater acoustic networks loss of connection and high bit error 
rate may be caused by shadow zones~\cite{Domingo2009} formed by different physical reasons. Domingo and Vuran (2012)~\cite{Domingo2012a} distinguish up to 5 different underwater 
propagation phenomena which may obstruct communication.

Faults can either be permanent or temporary~\cite{Henkel2011}. Permanent faults may be caused by manufacturing defect, as variances of the hardware components are 
inevitable due to physical reasons~\cite{Georgakos2013}. One of the other factors that can introduce faults is aging and wear-out of the hardware 
components~\cite{Lorenz2012}. In addition to the components themselves also the interconnection between them are affecting reliability and may cause faults~\cite{Sauli2012}. One of 
the challenges of fault management is temporary faults, especially soft errors. Soft Error is a temporary change of signal value due to ionizing particles~\cite{Henkel2011} that 
may lead to failure. Due to high integration density it is estimated that soft failure rate is increasing in future~\cite{Rehman2015}.

\subsection{Fault Tolerance Techniques}
A distributed system is defined~\cite{tanenbaum2007distributed} as a collection of independent computers that appears to its users as a single coherent system. A Sensor network 
consists of a number of sensor nodes that form the network and feed data to single or multiple data sinks. Provided that in the sensor network the sensor nodes are autonomous, it 
can be seen as a distributed system. Faults happening in sensor networks can be addressed using the same techniques as in distributed systems~\cite{tanenbaum2007distributed}.
The used techniques can be categorized into following groups:
\begin{itemize}
\item Fault Prediction and Prevention are about preventing a fault to happen and proactive fault avoidance.
\item Fault Detection and Identification are responsible for detecting and localizing of the fault.
\item Fault Isolation, Masking and Recovery are different techniques for repairing fault, minimizing the effect of a fault or avoiding it to turn to system failure.
\end{itemize}

% ===============================END OF INTRODUCTION===============================================
\section{Fault Prevention and Prediction}\label{sect:prevention}
Fault prevention and prediction in sensor networks is dependent on the initial deployment method of the sensor network and the architectural design of the system. These will be 
looked at in the following subsections.

\subsection{Design of the Sensor Network}
In Wireless Sensor Networks (WSN), instead of a centralized homogeneous topology, dividing nodes into clusters is an energy efficient and resilient method~\cite{Liu2012}, where 
dedicated cluster head nodes may have more energy and communication capabilities to effectively act as mediators between regular nodes and data sinks.

To overcome issues caused by varying environmental challenges of Underwater Wireless Sensor Networks (UWSN), natural algorithms may be utilized. For instance clustering and 
routing can be done utilizing Cuckoo Search algorithm and Particle Swarm Optimization~\cite{Sofi2018} which have behaved more resiliently in underwater conditions than more usual 
terrestrial Low Energy Adaptive Clustering Hierarchy (LEACH) protocol~\cite{Tyagi2013}. Pressure measurements have been used for UWSN routing~\cite{Noh2016} with floating 
depth-controlling sensors.
% Maarja: uurida rohkem wireless asju, sellel on suurem impact, seda tahame me võib-olla ise rohkem kasutada laboris
% more topologies, protocols?
Fault Management tasks can also be distributed across the whole network. In WSN with enough spare nodes energy efficient grid can be formed~\cite{Asim2008}, changing node manager, 
gateway and sensing nodes selected and spare nodes put to sleep. This results in energy-efficient and lightweight network but needs excess nodes.

However, existing UWSN protocols have not been adequately compared in underwater field trial yet~\cite{Jiang2018}.

\subsection{Sensor Network Deployment}
Sensor network deployment techniques are important for WSN where deployment may affect directly nodes' locations and networking availability. Even for 
terrestrial wireless sensor networks, to obtain a satisfactory network performance, an adaptable deployment method is essential~\cite{Wu2007a}. Usually sensor placement for 
WSN utilizes more sensors than the minimum required number for redundancy reasons~\cite{Isler2004}. The deployment costs and energy efficiency of WSN-s have been investigated 
 and it has been found that there is no single solution that can be easily applied in practice \cite{Cheng2008}.

Wired sensor network deployment is less researched, possibly because wired sensor networks' node deployment locations are limited to cable, their locations are more 
predetermined and node connectivity is not directly related to location.

\subsection{Data Collection}
Sensor networks tend to have limited network bandwidth, energy and storage capabilities, thus filtering and aggregating sensor information may be a way to meet requirements.
Raw sensor data near the source can be divided into informative, non-informative and outlier groups~\cite{Bhuvana2018} and only needed data communicated or stored. Outlier data 
may result from noise, failures, disturbances etc. and may be useful for fault tolerance purposes.

Different techniques to compress and aggregate collected information in UWSN-s are investigated~\cite{Goyal2017}. I was found that aggregation is justified and cluster-based 
aggregation techniques are performing better than non-cluster based or other.

Security challenges need to be addressed and one way to minimize the risk of data tampering and/or interference is to ensure that data is processed locally or if that is not 
possible then communicated end-to-end encrypted~\cite{Kohnstamm2014}.

\subsection{UWSN testing Frameworks}
\label{sect:frameworks}
% see osa on pildist väljaspool, on sellised management asjad, aga neid on vähe tehtud..
% lugeda neid ja vaadata, et miks need siin pealkirja all on, mis need tegelikult on?
For UWSN-s there have been developed DESERT framework version 1 and version 2~\cite{Campagnaro2016} and SUNSET framework~\cite{Petrioli2015} that 
allow simulation, emulation and testing of networks. A conducted analysis~\cite{Petroccia2013} shows that SUNSET represents a more mature, flexible and robust 
framework for in field testing than DESERT, but DESERT v2 was released after that. For acoustic UWSN security testing SecFUN framework~\cite{Ateniese2015} has been proposed.

\section{Fault Detection and Identification}\label{sect:detection}
In essence fault detection means determining that one or more bits in the computation differ from their correct value~\cite{Carter2010}. This can be detected via 
continuous  monitoring of the network and nodes' status. Some sources also use the word ``Diagnosis'' in a broader meaning than just detection and identification. 
Diagnosis has been defined as ``characterizing the system's state to locate the causes of errors, determine how the system is changing over time, and predict errors before they 
occur~\cite{Carter2010}``. Current section covers different techniques to execute previously mentioned concepts.

Distributed hierarchical fault management has been used~\cite{Liu2013} for WSN-s, where agent fault detection devices collect information from power modules and sensors to 
determine failure conditions and sequentially diagnose the nature of the detected failure. 

In industry on higher abstraction levels there has been wide use of SNMP~\cite{case1990simple} protocol for fault detection querying and triggering in IP networked devices. There 
are multiple commercial tools for generating failures, e.g Chaos Monkey from Netflix~\cite{Gunawi2011}, that randomly terminate services in production environments, to 
ensure resiliency of them. The latter do not  manage occurring faults but ensure that the repairing mechanisms are in place and operable. Intelligent Platform Management Interface 
(IPMI)~\cite{Intel2004} is an industrial technology specification for hardware system management and monitoring.

Neural-network-based scheme for sensor failure detection, identification, and accommodation can be used which may allow the conditions to deviate to
greater extent from theoretical models and estimation. Relatively simple and computationally light approach has been presented~\cite{Napolitano1995} where neural network is used 
as on-line learning state estimator for detecting faults. Neural  network itself can be built as fault-tolerant~\cite{Neti1992}, so that failing nodes have least impact on result 
data.

Situational Awareness approach, using a mechanism that has been borrowed from humans, can be used for Internet of Things (IoT) sensor data interpretation, specifically 
regarding processes of sensation, perception and cognition. In addition to specification-based and learning-based approaches, perception-based approach utilizing Fuzzy Formal 
Concept was proposed~\cite{Benincasa2014} for Situational Awareness identification.

Semantic Sensor Network Ontology has been proposed~\cite{Compton2012} for managing interoperability between sensing systems. The Semantic Ground describes information for 
interoperability and cooperation among agents~\cite{DAniello2016}. To enhance resilience in Semantic Sensor Networks, monitoring nodes may forward observations to association 
nodes, which develop Situational Awareness by mining association rules for example via natural Artificial Bee Colony algorithm~\cite{DAniello2016}.

Electrical Power Grids need efficient monitoring, for outage detection, environmental monitoring and fault diagnostics different WSN-based approaches are reviewed~\cite{Fadel2015}. 
Most of these approaches are also used in other applications.

% \subsection{Sensor Network Monitoring}
% \paragraph{Monitoring Node Status}
% \paragraph{Monitoring Congestion Level}
% \subsection{Recovery from composite Faults Including Congestion}
% \subsection{Composition of Fault Tolerance and Timeliness Requirements}
% \paragraph{Collection of Raw Data}
% \paragraph{Collection of Aggregated Data}
% \subsection{Reliable Data Dissemination}
% \paragraph{Hop-by-Hop Reliability}
% \paragraph{Minimum Dominating Set}
% \paragraph{Combining Source and Local Recovery}

\section{Fault Isolation, Masking and Recovery}\label{sect:recovery}
After fault detection, identification and diagnosing, fault handling stage can be entered~\cite{Liu2013} to prevent further data corruption and system deterioration. 
The fault handling stage consists of Fault Isolation, Masking and Recovery. Fault handling can hide the fault occurrence from other components - the key techniques for such 
masking are informational, time and physical redundancy~\cite{tanenbaum2007distributed}. Isolating a faulty component from others can be facilitated by using 
virtualization~\cite{tanenbaum2007distributed}. In large scale distributed systems frozen virtual images of healthy services have been used as checkpoints~\cite{Cristea2011} for 
rolling back in case of a fault occurrence.

Fault Recovery ensures that the fault does not propagate to visible results, for instance by rolling back to a previous healthy state (checkpointing) or re-trying failed 
operations (time redundancy). Some of the techniques for Fault Recovery can be Reconfiguration - changing the system's state so that the same or similar error is prevented from 
occurring again, and Adaptation - re-optimizing the system for instance after Reconfiguration task~\cite{Carter2010}.

In Sensor Networks, different approaches for Fault Recovery have been used, that have different resource overheads, energy-efficiencies, scalabilities and 
network types. For both network and node fault recovery in  wireless sensor networks Mitra (2016)~\cite{Mitra2016} compares checkpoint based (CRAFT), agent based 
recovery (ABSR),  fault node recovery algorithm (FNR), cluster-based and hierarchical fault management (CHFM), Failure Node Detection and Recovery algorithm (FNDRA). While some of
those are specific for terrestrial wireless usage, some principles (e.g checkpointing etc.) can also be used in wired and/or underwater environments. To reduce network bandwidth 
requirements checkpoint backup can be mobile to nearby nodes~\cite{Salera2007} and used for recovering from fault situations.

In networks, error control schemes are commonly classified into three groups~\cite{Domingo2012a}:
\begin{itemize}
\item Automatic Repeat Request (ARQ) - retransmission of corrupted data is asked
\item Forward Error Correction (FEC) - data corruption can be detected and corrected by receiving end
\item Hybrid ARQ (HARQ) - combination of FEC and ARQ
\end{itemize}
These groups are similar to already mentioned techniques for node fault management. 

Cross-layer approach benefits fault recovery significantly because single layer redundancy, such as hardware redundancy and application checkpointing, have very high costs and 
a wide variance in delay between fault occurrence and detection makes recovery difficult~\cite{Carter2010}.

\section{Open Research Issues}
\label{sect:open}
\subsection{Security}
Faults and security are interrelated concepts~\cite{Cristea2011}. It needs effort to prevent systems from being penetrated when working as intended, faults add uncertainty and 
make the task of prevention even  harder. Faults can be created by an intrusion, but moreover faults can enable new intrusion vectors~\cite{DeHon2010} - misbehaving devices 
violate key assumptions and create number of new attack vectors to systems. For example soft errors explained in section~\ref{sect:sources} can be used to defeat 
cryptography~\cite{Xu2001}.

\subsection{Energy-efficiency}
Power dissipation has now reached a point where energy concerns limit the computation we can deploy on chip~\cite{DeHon2010} and the aim is shifting from transistor density and 
speed to energy density and cost. Energy density and efficiency needs to be addressed also on larger-scale, for instance WSN-s may not have unlimited power supply and 
need to utilize energy-efficiency strategies~\cite{Cheng2008,Asim2008, Liu2012,Tyagi2013}. For fault tolerance techniques, cross-layer approach is considered more energy-efficient 
~\cite{Carter2010} than single layer. Strategic redundancy in cross-layer approach may allow systems to safely operate on the verge of failure~\cite{DeHon2010} spending less 
energy without going over the edge.

\subsection{Scalability}
One of traditional benefits of scaling has been the decrease of cost per functionality~\cite{DeHon2010}, but easing reliability problems by multiplicating logic, voting and 
similar techniques means that the scaled technology might not offer a reduction of energy or area. Some tolerance techniques may increase computing overhead, and not all 
approaches are scalable~\cite{Mitra2016}. Large scale fault tolerant systems are researched~\cite{Cristea2011} without paying special attention to energy and communication 
constraints.

\subsection{Cross Layer Approach to Fault Tolerance}
Faults are not going to disappear but likely to increase in future~\cite{Rehman2015}. One way to cope with faults is to accept imperfect devices to fail and compensate failures at 
higher levels in system stack~\cite{DeHon2010}, tolerating faults cross layer involving circuit design, firmware, operating system, applications etc. Cross-layer fault tolerant 
systems have potential to implement reliable, high-performance and energy-efficient solutions without overwhelming cost~\cite{Carter2010} by distributing the responsibilities of 
tolerating faults across multiple layers~\cite{Veleski}. 

In case fault detection and fault recovery are to be implemented in different system layers then following challenges arise~\cite{Mitra2010a}:
\begin{itemize}
\item For statistical validation and metrics high confidence resource-light reliability and availability estimation is needed.
\item Verification of resilience techniques, to be sure that resilience techniques perform under all possible scenarios.
\item Reliability grades for testing and grading system-wide reliability and data integrity. Reliability may change under workload.
\end{itemize}

In addition to Cross Layer approach also Multi Layer approach~\cite{Henkel2014a} has been proposed, where system layers are adapted to each other to reduce error 
propagation. In opinion of the author of current paper, this is not a separate approach, but rather a small increment of Cross Layer approach.

\subsection{Specifics of USN networks}
Underwater environment is mostly different because of harsh physical conditions - pressure, hard accessibility, limited communication and energy resources. Many communication 
methods are unavailable underwater and there are multiple phenomena~\cite{Domingo2009,Domingo2012a} that obstruct communication. Because of the possibility of flooding hardware, 
more attention and resources should be paid to physical security. On the other hand faults from excessive heat should be rare and avoidable underwater.

While most of the common concepts should be possible to be adapted for underwater use, the environment is more demanding and unforgiving and faults 
are more costly. Some more demanding approaches like cloud computing may not make sense to implement in USN, but author cannot see any low network bandwidth and power 
requirement fault tolerant approach mentioned in current paper, that cannot be used underwater. One of the more promising approaches that could be adapted well between 
underwater environment's constraints seems to be cross-layer resilience, which for unknown reasons is lacking recent research papers even for terrestrial implementations.

\section{Conclusion}
\label{sect:conclusion}
Current paper presented fault tolerant techniques, presented a survey on fault tolerant techniques in USN-s  and pointed out open research issues in this field.
Fault tolerance is addressed in underwater context for reliant UWSN networking~\cite{Xu2012, Zenia2016, Lal2016, Domingo2012a}, space localization 
\cite{Das2017} and monitoring underwater pipelines~\cite{Mohamed2011}. Current paper overviewed fault tolerant techniques that are developed for underwater use or could be adapted 
for that. The techniques were divided into groups that are used in distributed systems and papers utilizing the techniques discussed in corresponding sections. 

Current paper is the first to investigate fault tolerance, particularly cross layer fault tolerance, in USN-s. According to the survey there is no research covering 
cross-layer fault tolerance for underwater sensor networks.

% Conclusion ära muuta, kuivemaks, mainida et oli esimene töö mis seda, mis tehtud sai
% contribution: esimest korda sellel teemal, open research issued sellest valdkonnas

\bibliographystyle{ieeetr}
\bibliography{ftt_in_usn}

\begin{thebibliography}{10}

\bibitem{Kao2017}
C.-C. Kao, Y.-S. Lin, G.-D. Wu, and C.-J. Huang, ``{A study of applications,
  challenges, and channel models on the Internet of Underwater Things},'' in
  {\em {2017 Int. Conf. Appl. Syst. Innov.}}, no.~2, pp.~1375--1378, IEEE, may
  2017.

\bibitem{Khosrow-PourD.B.A.2018}
S.~Kumar and {Balamurugan B}, ``{Fault Tolerant Cloud Systems},'' in {\em
  {Encycl. Inf. Sci. Technol. Fourth Ed.}} (M.~{Khosrow-Pour, D.B.A.}, ed.),
  ch.~93, pp.~1075--1090, IGI Global, 4th~ed., 2018.

\bibitem{Wilfredo2000}
T.~Wilfredo and W.~Torres-Pomales, ``{Software Fault Tolerance: A Tutorial},''
  Tech. Rep. October, NASA, 2000.

\bibitem{Jaynes2013}
E.~Jaynes and F.~Cummings, ``{Fault Management in Wireless Sensor Networks},''
  {\em Proc. IEEE}, vol.~51, no.~1, pp.~89 -- 109, 2013.

\bibitem{Paradis2007}
L.~Paradis and Q.~Han, ``{A survey of fault management in wireless sensor
  networks},'' {\em J. Netw. Syst. Manag.}, vol.~15, no.~2, pp.~171--190, 2007.

\bibitem{Domingo2009}
M.~C. Domingo, ``{A topology reorganization scheme for reliable communication
  in underwater wireless sensor networks affected by shadow zones},'' {\em
  Sensors}, vol.~9, no.~11, pp.~8684--8708, 2009.

\bibitem{Domingo2012a}
M.~C. Domingo and M.~C. Vuran, ``{Cross-layer analysis of error control in
  underwater wireless sensor networks},'' {\em Comput. Commun.}, vol.~35,
  no.~17, pp.~2162--2172, 2012.

\bibitem{Henkel2011}
J.~Henkel, L.~Hedrich, A.~Herkersdorf, R.~Kapitza, D.~Lohmann, P.~Marwedel,
  M.~Platzner, W.~Rosenstiel, U.~Schlichtmann, O.~Spinczyk, M.~Tahoori,
  L.~Bauer, J.~Teich, N.~Wehn, H.-J. Wunderlich, J.~Becker, O.~Bringmann,
  U.~Brinkschulte, S.~Chakraborty, M.~Engel, R.~Ernst, and H.~H{\"a}rtig,
  ``{Design and architectures for dependable embedded systems},'' in {\em
  {Proc. seventh IEEE/ACM/IFIP Int. Conf. Hardware/software codesign Syst.
  Synth. - CODES+ISSS '11}}, (New York, New York, USA), p.~69, ACM Press, 2011.

\bibitem{Georgakos2013}
G.~Georgakos, U.~Schlichtmann, R.~Schneider, and S.~Chakraborty, ``{Reliability
  challenges for electric vehicles},'' in {\em {Proc. 50th Annu. Des. Autom.
  Conf. - DAC '13}}, (New York, New York, USA), p.~1, ACM Press, 2013.

\bibitem{Lorenz2012}
D.~Lorenz, M.~Barke, and U.~Schlichtmann, ``{Efficiently analyzing the impact
  of aging effects on large integrated circuits},'' {\em Microelectron.
  Reliab.}, vol.~52, pp.~1546--1552, aug 2012.

\bibitem{Sauli2012}
Z.~Sauli, V.~Retnasamy, S.~Taniselass, A.~H. Shapri, R.~M. Hatta, and M.~H.
  Aziz, ``{Polymer core BGA vertical stress loading analysis},'' {\em Proc.
  Int. Conf. Comput. Intell. Model. Simul.}, vol.~129, pp.~148--151, 2012.

\bibitem{Rehman2015}
S.~Rehman, {\em {Reliable Software for Unreliable Hardware -- A Cross-Layer
  Approach}}.
\newblock Doctoral dissertation, Karlsruhe Institute of Technology (KIT), 2015.

\bibitem{tanenbaum2007distributed}
A.~S. Tanenbaum and M.~{Van Steen}, {\em {Distributed systems: principles and
  paradigms}}.
\newblock Prentice-Hall, 2007.

\bibitem{Liu2012}
S.~P. Singh and S.~C. Sharma, ``{A survey on cluster based routing protocols in
  wireless sensor networks},'' {\em Procedia Comput. Sci.}, vol.~45, no.~C,
  pp.~687--695, 2015.

\bibitem{Sofi2018}
S.~A. Sofi and R.~N. Mir, ``{Natural algorithm based adaptive architecture for
  underwater wireless sensor networks},'' {\em Proc. 2017 Int. Conf. Wirel.
  Commun. Signal Process. Networking, WiSPNET 2017}, vol.~2018-Janua,
  pp.~2343--2346, 2018.

\bibitem{Tyagi2013}
S.~Tyagi and N.~Kumar, ``{A systematic review on clustering and routing
  techniques based upon LEACH protocol for wireless sensor networks},'' {\em J.
  Netw. Comput. Appl.}, vol.~36, no.~2, pp.~623--645, 2013.

\bibitem{Noh2016}
Y.~Noh, U.~Lee, S.~Lee, P.~Wang, L.~F. Vieira, J.~H. Cui, M.~Gerla, and K.~Kim,
  ``{HydroCast: Pressure routing for underwater sensor networks},'' {\em IEEE
  Trans. Veh. Technol.}, vol.~65, no.~1, pp.~333--347, 2016.

\bibitem{Asim2008}
M.~Asim, H.~Mokhtar, and M.~Merabti, ``{A fault management architecture for
  wireless sensor network},'' {\em IWCMC 2008 - Int. Wirel. Commun. Mob.
  Comput. Conf.}, pp.~779--785, 2008.

\bibitem{Jiang2018}
S.~Jiang, ``{State-of-the-Art Medium Access Control (MAC) Protocols for
  Underwater Acoustic Networks: A Survey Based on a MAC Reference Model},''
  {\em IEEE Commun. Surv. Tutorials}, vol.~20, no.~1, pp.~96--131, 2018.

\bibitem{Wu2007a}
C.~H. Wu, K.~C. Lee, and Y.~C. Chung, ``{A Delaunay Triangulation based method
  for wireless sensor network deployment},'' {\em Comput. Commun.}, vol.~30,
  no.~14-15, pp.~2744--2752, 2007.

\bibitem{Isler2004}
V.~Isler, S.~Kannan, and K.~Daniilidis, ``{Sampling Based Sensor-Network
  Deployment},'' pp.~1780--1785, 2004.

\bibitem{Cheng2008}
Z.~Cheng, M.~Perillo, and W.~B. Heinzelman, ``{General network lifetime and
  cost models for evaluating sensor network deployment strategies},'' {\em IEEE
  Trans. Mob. Comput.}, vol.~7, no.~4, pp.~484--497, 2008.

\bibitem{Bhuvana2018}
V.~P. Bhuvana, C.~Preissl, A.~M. Tonello, and M.~Huemer, ``{Multi-Sensor
  Information Filtering with Information-Based Sensor Selection and Outlier
  Rejection},'' {\em IEEE Sens. J.}, vol.~18, pp.~2442--2452, mar 2018.

\bibitem{Goyal2017}
N.~Goyal, M.~Dave, and A.~K. Verma, ``{Data aggregation in underwater wireless
  sensor network: Recent approaches and issues},'' {\em J. King Saud Univ. -
  Comput. Inf. Sci.}, 2017.

\bibitem{Kohnstamm2014}
J.~Kohnstamm and D.~Madhub, ``{Mauritius Declaration},'' pp.~1--2, 2014.

\bibitem{Campagnaro2016}
F.~Campagnaro, R.~Francescon, F.~Guerra, F.~Favaro, P.~Casari, R.~Diamant, and
  M.~Zorzi, ``{The DESERT underwater framework v2: Improved capabilities and
  extension tools},'' {\em 3rd Underw. Commun. Netw. Conf. Ucomms 2016}, 2016.

\bibitem{Petrioli2015}
C.~Petrioli, R.~Petroccia, J.~R. Potter, and D.~Spaccini, ``{The SUNSET
  framework for simulation, emulation and at-sea testing of underwater wireless
  sensor networks},'' {\em Ad Hoc Networks}, vol.~34, pp.~224--238, 2015.

\bibitem{Petroccia2013}
R.~Petroccia and D.~Spaccini, ``{Comparing the SUNSET and DESERT frameworks for
  in field experiments in underwater acoustic networks},'' {\em Ocean. 2013
  MTS/IEEE Bergen Challenges North. Dimens.}, 2013.

\bibitem{Ateniese2015}
G.~Ateniese, A.~Capossele, P.~Gjanci, C.~Petrioli, and D.~Spaccini, ``{SecFUN:
  Security framework for underwater acoustic sensor networks},'' {\em MTS/IEEE
  Ocean. 2015 - Genova Discov. Sustain. Ocean Energy a New World}, 2015.

\bibitem{Carter2010}
N.~P. Carter, H.~Naeimi, and D.~S. Gardner, ``{Design techniques for
  cross-layer resilience},'' {\em 2010 Des. Autom. Test Eur. Conf. Exhib. (DATE
  2010)}, no.~February, pp.~1023--1028, 2010.

\bibitem{Liu2013}
T.-H. Liu, S.-C. Yi, and X.-W. Wang, ``{A fault management protocol for
  low-energy and efficient Wireless sensor networks},'' {\em J. Inf. Hiding
  Multimed. Signal Process.}, vol.~4, no.~1, pp.~34--45, 2013.

\bibitem{case1990simple}
J.~D. Case, M.~Fedor, M.~L. Schoffstall, and J.~Davin, ``{Simple network
  management protocol (SNMP)},'' tech. rep., 1990.

\bibitem{Gunawi2011}
H.~S. Gunawi, T.~Do, J.~M. Hellerstein, I.~Stoica, D.~Borthakur, and
  J.~Robbins, ``{Failure as a service (faas): A cloud service for large-scale,
  online failure drills},'' {\em Electr. Eng. Comput. Sci.}, pp.~1--8, 2011.

\bibitem{Intel2004}
Intel, ``{Intelligent Platform Management Interface},'' 2004.

\bibitem{Napolitano1995}
M.~R. Napolitano, C.~Neppach, V.~Casdorph, S.~Naylor, M.~Innocenti, and
  G.~Silvestri, ``{Neural-network-based scheme for sensor failure detection,
  identification, and accommodation},'' {\em J. Guid. Control. Dyn.}, vol.~18,
  pp.~1280--1286, nov 1995.

\bibitem{Neti1992}
C.~Neti, M.~H. Schneider, and E.~D. Young, ``{Maximally Fault Tolerant Neural
  Networks},'' {\em IEEE Trans. Neural Networks}, vol.~3, no.~1, pp.~14--23,
  1992.

\bibitem{Benincasa2014}
G.~Benincasa, G.~D'Aniello, C.~{De Maio}, V.~Loia, and F.~Orciuoli, ``{Towards
  perception-oriented situation awareness systems},'' {\em Adv. Intell. Syst.
  Comput.}, vol.~322, pp.~813--824, 2014.

\bibitem{Compton2012}
M.~Compton, P.~Barnaghi, L.~Bermudez, R.~Garc{\'i}a-Castro, O.~Corcho, S.~Cox,
  J.~Graybeal, M.~Hauswirth, C.~Henson, A.~Herzog, V.~Huang, K.~Janowicz, W.~D.
  Kelsey, D.~{Le Phuoc}, L.~Lefort, M.~Leggieri, H.~Neuhaus, A.~Nikolov,
  K.~Page, A.~Passant, A.~Sheth, and K.~Taylor, ``{The SSN ontology of the W3C
  semantic sensor network incubator group},'' {\em J. Web Semant.}, vol.~17,
  pp.~25--32, 2012.

\bibitem{DAniello2016}
G.~D'Aniello, A.~Gaeta, and F.~Orciuoli, ``{Artificial bees for improving
  resilience in a sensor middleware for Situational Awareness},'' {\em TAAI
  2015 - 2015 Conf. Technol. Appl. Artif. Intell.}, pp.~300--307, 2016.

\bibitem{Fadel2015}
E.~Fadel, V.~C. Gungor, L.~Nassef, N.~Akkari, M.~G. {Abbas Malik}, S.~Almasri,
  and I.~F. Akyildiz, ``{A survey on wireless sensor networks for smart
  grid},'' {\em Comput. Commun.}, vol.~71, pp.~22--33, 2015.

\bibitem{Cristea2011}
V.~Cristea, C.~Dobre, F.~Pop, C.~Stratan, A.~Costan, C.~Leordeanu, and
  E.~Tirsa, ``{A dependability layer for large-scale distributed systems},''
  {\em Int. J. Grid Util. Comput.}, vol.~2, no.~2, p.~109, 2011.

\bibitem{Mitra2016}
S.~Mitra, ``{Comparative Study of Fault Recovery Techniques in Wireless Sensor
  Network},'' no.~December, pp.~19--21, 2016.

\bibitem{Salera2007}
I.~Salera, A.~Agbaria, and M.~Eltoweissy, ``{Fault-tolerant mobile sink in
  networked sensor systems},'' {\em 2006 2nd IEEE Work. Wirel. Mesh Networks,
  WiMESH 2006}, pp.~106--108, 2007.

\bibitem{DeHon2010}
A.~DeHon, H.~M. Quinn, and N.~P. Carter, ``{Vision for cross-layer optimization
  to address the dual challenges of energy and reliability},'' {\em Proc.
  -Design, Autom. Test Eur. DATE}, no.~March, pp.~1017--1022, 2010.

\bibitem{Xu2001}
J.~Xu, S.~Chen, Z.~Kalbarczyk, and R.~K. Iyer, ``{An experimental study of
  security vulnerabilities caused by errors},'' {\em Proc. Int. Conf.
  Dependable Syst. Networks}, pp.~421--430, 2001.

\bibitem{Veleski}
M.~Veleski, R.~Kraemer, and M.~Krstic, ``{An Overview of Cross-Layer Resilience
  Design Methods},'' in {\em {Fifth Int. Conf. Radiat. Appl. Var. Fields
  Res.}}, 2017.

\bibitem{Mitra2010a}
S.~Mitra, K.~Brelsford, and P.~N. Sanda, ``{Cross-layer resilience challenges:
  Metrics and optimization},'' {\em 2010 Des. Autom. Test Eur. Conf. Exhib.
  (DATE 2010)}, no.~March 2010, pp.~1029--1034, 2010.

\bibitem{Henkel2014a}
J.~Henkel, L.~Bauer, H.~Zhang, S.~Rehman, and M.~Shafique, ``{Multi-Layer
  Dependability: From Microarchitecture to Application Level},'' {\em Proc.
  51st Annu. Des. Autom. Conf. Des. Autom. Conf.}, pp.~47:1--47:6, 2014.

\bibitem{Xu2012}
J.~Xu, K.~Li, and G.~Min, ``{Reliable and energy-efficient multipath
  communications in underwater sensor networks},'' {\em IEEE Trans. Parallel
  Distrib. Syst.}, vol.~23, no.~7, pp.~1326--1335, 2012.

\bibitem{Zenia2016}
N.~Z. Zenia, M.~Aseeri, M.~R. Ahmed, Z.~I. Chowdhury, and M.~{Shamim Kaiser},
  ``{Energy-efficiency and reliability in MAC and routing protocols for
  underwater wireless sensor network: A survey},'' {\em J. Netw. Comput.
  Appl.}, vol.~71, pp.~72--85, 2016.

\bibitem{Lal2016}
C.~Lal, R.~Petroccia, M.~Conti, and J.~Alves, ``{Secure underwater acoustic
  networks: Current and future research directions},'' {\em 3rd Underw. Commun.
  Netw. Conf. Ucomms 2016}, 2016.

\bibitem{Das2017}
A.~P. Das and S.~M. Thampi, ``{Fault-resilient localization for underwater
  sensor networks},'' {\em Ad Hoc Networks}, vol.~55, pp.~132--142, 2017.

\bibitem{Mohamed2011}
N.~Mohamed, I.~Jawhar, J.~Al-Jaroodi, and L.~Zhang, ``{Sensor network
  architectures for monitoring underwater pipelines},'' {\em Sensors}, vol.~11,
  no.~11, pp.~10738--10764, 2011.

\end{thebibliography}
\vspace{12pt}
\end{document}